# Strong influence of packing density in terahertz metamaterials

Ranjan Singh,[1,2,a] Carsten Rockstuhl,[3] and Weili Zhang[1]

[1]School of Electrical and Computer Engineering,
Oklahoma State University, Stillwater, Oklahoma 74078, USA

[2]Center for Integrated Nanotechnologies, Materials Physics and Applications Division, Los Alamos National Laboratory, Los Alamos, New Mexico 87545, USA

[3] Institute of Condensed Matter Theory and Solid State Optics, Friedrich-Schiller-Universität Jena, Jena 07743, Germany

**Abstract**

We investigate the response of terahertz metamaterials made of split-ring resonators depending on the unit cell density. It is shown that the fundamental resonance has its highest $Q$ factor for a period of one-third the resonance wavelength. Increasing or decreasing the period from that optimal period reduces the $Q$ factor. All observations are explained by understanding radiative coupling as the dominant loss mechanism at low densities, whereas the strong inter SRR coupling enhances non-radiative losses at high densities. Our results allow designing metamaterials with adjustable $Q$ factors and possibly suggesting that an optimal package density exists to induce the strongest dispersion.

[a]Electronic mail: ranjan@lanl.gov

Metamaterials (MMs) constitute an exploding field of science and technology. It allows to mold the flow of electromagnetic waves almost arbitrarily using suitably tailored unit cells. Potentially the most common unit cell for MMs is the split ring resonator (SRR) **[1]**. The sub-wavelength closely packed SRR allows accessing several functionalities such as optical magnetism, negative refraction, frequency selection, modulation, and sensing capabilities **[2-6]**. The impact on the light propagation depends strongly on the illumination and polarization direction with respect to the SRR. The unifying aspect in all such *modi operandi* is that the properties of the MM are predominantly determined by the resonant scattering response of light at individual SRRs. It is therefore essential to come up with schemes to maximize the quality ($Q$) factor of the SRR resonances in order to optimize their performances for various applications. Moreover, the stronger the resonance the stronger the obtainable dispersion in any effective parameter describing conceptually the MM. Since the $Q$ factor is limited by either radiative or non-radiative losses, their reduction is usually understood as to be essential to achieve resonances with high $Q$ factors.

The $Q$ factor in symmetrical SRRs usually suffers from a strong coupling to free space radiation, resulting in high radiative losses. To overcome such problem, Fedotov *et al.* demonstrated that high $Q$ factors can be achieved by relying on dark modes **[7]**. They are sustained by SRRs which are only weakly coupled to free space by a gentle break in symmetry **[8,9]**. Non-radiative losses can be affected by controlling the distance between individual SRRs or cooling them to cryogenic temperatures to exploit super-conductive properties of involved materials. A third aspect which affects the $Q$ factor is radiative interaction of SRRs in the lattice. Such interaction was demonstrated in previous works by investigating the impact of disorder in the unit cell arrangement **[10-12]** and the impact of laterally coupled SRRs **[13,14]** on the resonance. Bitzer *et al.* showed that the inter SRR coupling is mediated by lattice modes and it can have influence on the resonance sharpness [15]. In another recent work Fedotov *et al.* observed the collapse of a sharp Fano resonance with an increasing number of unit cell resonators but keeping the distance between the periodic metamolecules fixed [16]. All these investigations point towards the importance of suppressing radiative coupling among adjacent SRRs to enhance their $Q$ factors. The decrease of the $Q$ factors can be understood by considering the driving field of each SRR in the lattice. On the one hand, the SRR is driven by the external illumination. On the other hand, it is driven by the scattered field from neighboring SRRs. If the distance between neighboring SRRs is too large, both fields tend to be out of phase, causing a decrease of the driving force since they interfere destructively. In consequence, the SRRs should be placed as close as possible to suppress such coherent interactions. On the other hand, placing them too narrow favors non-radiative losses by capacitive and inductive coupling. Therefore, to obtain some final conclusions it is compulsory to investigate whether increasing or decreasing the number density of the SRR in the lattice has an effect on the quality factor of the fundamental resonance and to find an answer on what could be an optimal period.

In this Letter, we demonstrate that the inter SRR distance and its number density in a fixed area can control the $Q$ factor and the strength of the inductive capacitive (LC) resonance of SRRs. We observe a strong suppression of radiative losses and thus a sharp enhancement of the $Q$ factor at a certain critical periodicity. When the periodicity is further reduced a gradual decline in $Q$ factor of the fundamental SRR resonance is observed. The minimal transmission in resonance shows an exponential decay as a function of the number density. Tailoring the periodicity of MMs will allow the design of frequency selective surfaces with an adjustable $Q$ factor for various applications and for optimized MMs across the entire electromagnetic spectrum. This finding will eventually lead to the development of metamolecule density tuned terahertz devices and components.

In the experiments we rely on a typical 8*f* confocal terahertz time-domain spectroscopy (THz-TDS) system consisting of a terahertz transmitter, beam collecting and steering optics, and a terahertz receiver **[17,18]**. The transmitter is a photoconducting dipole antenna fabricated on a GaAs chip. This antenna is dc biased at 70 V, and when it is excited by 25 fs optical pulses with a central wavelength of 800 nm and a repetition rate of 88 MHz from a Ti:sapphire laser, terahertz pulses are generated with a bandwidth of 0.2-4.5 THz. The collected terahertz pulse has a frequency-independent beam waist located at the focal position. The receiver chip is gated by the sampling laser pulse split off from the same laser beam as the excitation laser pulse for the transmitter. By changing the time delay between the excitation and detection laser beams, one can obtain a terahertz time-domain pulse trace both in amplitude and phase. The diameter of the terahertz beam is 3.5 mm at the focus and the S/N ratio is about 15000:1 for a single scan.

The MM is formed by a planar array of subwavelength SRRs with different periods as shown in Figs. 1(a) and 1(b). The SRR metal film pattern was fabricated using photolithographic technique and then depositing 200 nm of Aluminum metal on 640 µm thick n-type silicon substrate ($\varepsilon$ = 11.68). The geometrical dimensions of the SRR are shown in Fig. 1(c). Retaining the identical structure of the SRR, nine sets of samples were prepared with periods of *a* = 200, 150, 100, 60, 50, 40, 35, 30, and 25 µm. Figures 1(a) and 1(b) show the MM sample with *a* = 200 and 25 µm, respectively. The size of SRR was chosen such that its resonance wavelength is $\lambda$ = 200 µm. For samples with different periods, the resonance wavelength to lattice constant ratio $\lambda/a$ varies; ranging from a value of $\lambda/a$ =1 for *a* = 200 µm to $\lambda/a = 8$ for $a = 25$ μm. Each MM sample is excited by a incident terahertz beam 3.5 mm in diameter. Therefore, in the most diluted SRR sample with $a = 200$ μm, the number of SRRs being excited is around 960 and in the sample with the highest density where $a = 25$ μm, the number of SRRs being excited is 61,500. For a fixed area of terahertz beam excitation ($\frac{\pi}{4}(3.5mm)^2$), the density of SRRs is gradually increased from 960 to 61,500 in order to probe the inter SRR coupling and its impact on the overall resonant response of terahertz transmission. The MM sample size was kept fixed at 10 mm × 10 mm and the incident terahertz field was polarized with the electric field parallel to the SRR gap at normal incidence [see Fig. 1(c)]. A blank silicon wafer with no pattern is used as a reference.

Figure 2 shows the measured time domain transmission terahertz pulses through the samples with periodicity, $a = 100, 60, 40$ and $25$ μm as they were illuminated at normal incidence exciting the fundamental LC resonance. We observe significant reshaping of the terahertz pulse for different periodicities. A noteworthy observation is the high oscillation in the pulse for $a = 60$ μm. The transmission spectra in Fig. 3(a) are obtained by normalizing the measured transmission to the reference transmission of a blank n-type silicon wafer identical to the sample substrate for all the nine sets of MM sample with different periodicities. It is defined as $|E_s(\omega)/E_r(\omega)|$, where $E_s(\omega)$ and $E_r(\omega)$ are Fourier transformed time traces of the transmitted electric fields of the signal and the reference pulses, respectively. With the incident terahertz field polarization being along the SRR gaps, the LC resonance of the MM is excited due to charge distribution asymmetry which leads to the flow of circular current in the SRR loop. The LC resonance is extremely important in manipulating the dispersion in the effective permittivity at normal incidence as well as the effective permeability at an in-plane illumination. Figure 3(a) shows the measured spectra for MM samples with different periodicities. The evolution of the LC resonance can be observed at around 1.5 THz. The resonance was fairly weak for the 200, 150 and 100 μm periodicities. A strong resonance feature is observed for a MM period of 60 μm and then a gradual broadening of the resonance is seen for lower periodicities or higher packing densities. Simulations of the experiment based on the Fourier Modal Method reproduce the transmission measurement and are shown in Fig. 3(b). The 60 μm periodicity also corresponds to a λ/a ratio of 3.33 and this condition gives rise to the strongest resonance. We define this MM to be an *optimally packed* MM. The samples with higher periodicity than 60 μm or with (λ/a) < 3.33 can be termed as *loosely packed* MM and the ones with lower periodicity or (λ/a) > 3.33 can be defined as *tightly packed* MM.

The measured transmission resonance minimum is plotted in Fig. 4(a). We observe that the transmission at resonance decreases exponentially with the increasing density of metamolecules. The periodicity corresponding to the number of metamaterial unit cells is plotted on the right hand *y* axis of the same graph. Figure 4(b) indicates the strong dependence of the *Q* factor on the density of unit cells. Although, especially for the MM with the low packing density, it is slightly difficult to evaluate the *Q* factor very precisely, it can be seen that the *Q* factor rises when the MM array reaches the optimum density of SRRs and then gradually declines with a further increasing density. The lower *Q* factor for low densities is associated to an enhanced coupling to free-space radiation; especially into the silicon substrate where the index is high. Moreover, the strong radiative coupling among neighboring SRRs potentially causes the weakening of the resonance and also the slightly asymmetric line shape, e.g. for the sample with a period of 60 μm both in the experiment and the simulation. The collapse of the *Q* factor for small periods is mainly due the increase in non-radiative damping when the metamolecule density becomes high and the inter-particle coupling is strong.

The LC resonance in the SRR is excited by the incident electric field along the SRR gap side, giving rise to magnetic dipole moments in the direction perpendicular to the plane of SRR array when the incident wave illuminates the metamaterial at normal incidence. As we increase the density of SRR array starting from the resonance wavelength of an individual SRR, we observe the evolution of the LC resonance with the resonance getting sharpest at a periodicity of $\lambda/3.33$. As we further decrease the SRR periodicity beyond that of optimal packing, we observe a blue shift as well as broadening of the LC resonance. The blue shifting of the resonance is potentially due to the near field transverse magnetic dipole coupling between individual SRRs. The broadening is due to increase in non-radiative damping that effectively reduces the cross section per SRR.

In summary, we have experimentally and numerically demonstrated that the metamaterial arrays have the best resonance response for an optimum lattice periodicity. Tightly packed metamaterials favor non-radiative damping, leading to broadening of the fundamental LC resonance and thus a decline in the $Q$ factor. The dependence of the $Q$ factor on the distance between the SRRs provides a method for designing sub-wavelength terahertz metamaterial cavities with an adjustable $Q$ factor.

The authors acknowledge fruitful discussions with D. R. Chowdhury, M. T. Reiten, and J. Zhou. This work was partially supported by the US National Science Foundation, and the German Federal Ministry of Education and Research (Metamat) and the DAAD within the PPP program.

**Figure Captions**

FIG. 1 Microscopic image of the metamaterial array with lowest packing density of split ring resonators and periodicity $P = 200$ μm (a) and with highest packing density and $P = 25$ μm in (b). (c) unit cell with dimensions. The metafilm thickness = 200 nm for all metamaterial arrays.

FIG. 2 (color online) Measured sub-picosecond transmitted pulses through the metamaterial with different periodicity. The inset shows the magnified oscillations in the pulses at later times. Incident E field is polarized parallel to the gap of the double SRRs.

FIG. 3 (color online). (a) Measured and (b) simulated transmission spectra of varying periodicity metamaterial.

FIG. 4 (color online). (a) Percentage transmission at resonance and (b) extracted $Q$-factor from experimental data for the different number of metamolecules excited by the terahertz wave; Blue and red curves are just to guide the eye and black curve is an exponential fit to the transmission.

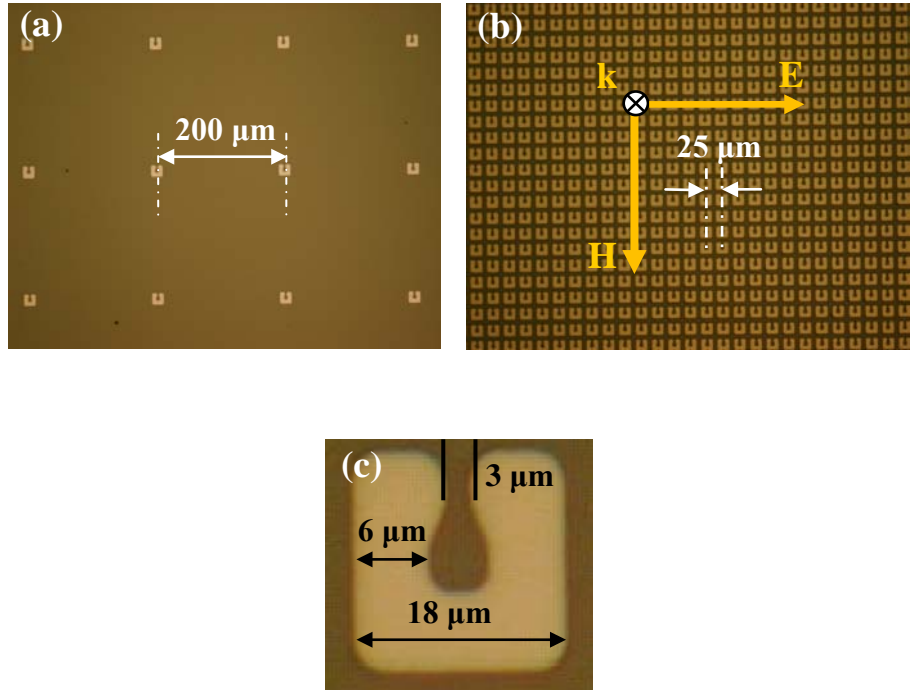

**Figure 1.
Singh** *et al.*

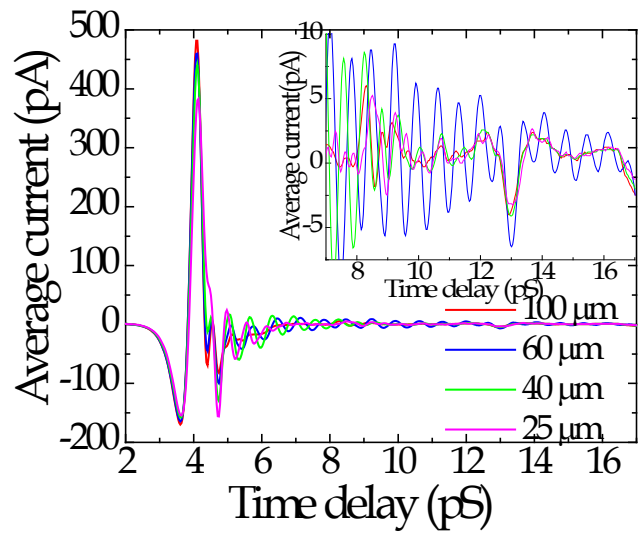

**Figure 2.**
**Singh** *et al.*

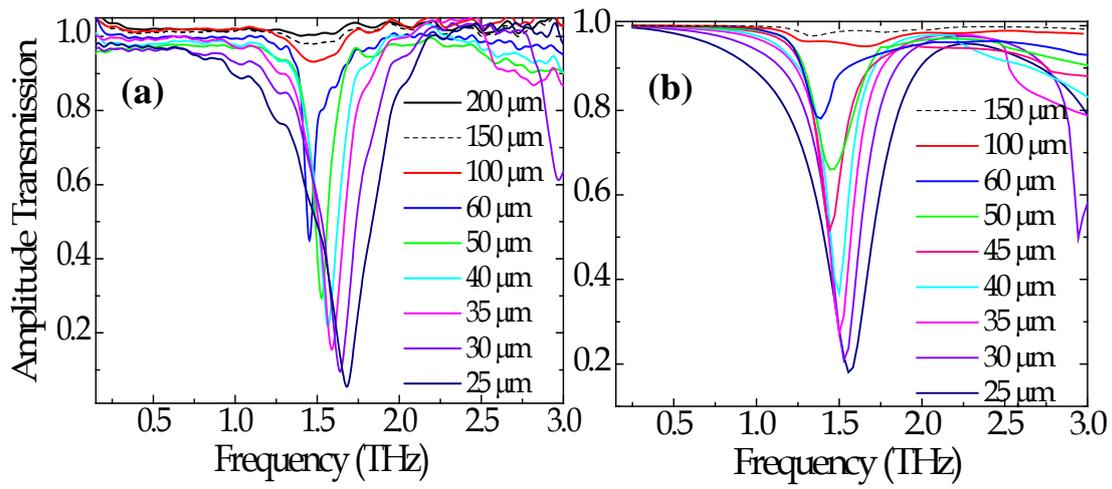

**Figure 3.
Singh *et al.***

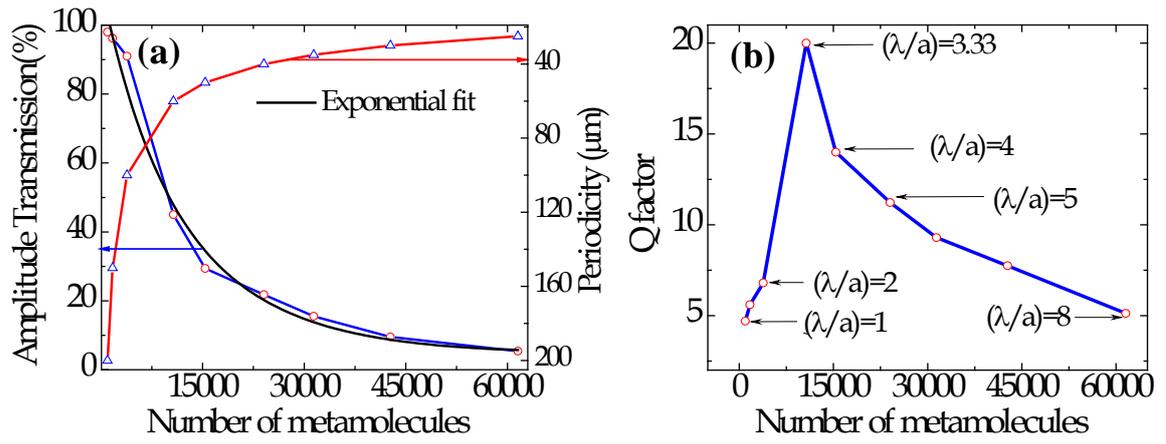

**Figure 4.**
**Singh *et al.***